\documentclass{sig-alternate}

\usepackage{amsmath}
\usepackage{amssymb}
\usepackage{algorithm}
\usepackage{todonotes}
\usepackage[noend]{algorithmic}

\begin{document}
%
\conferenceinfo{XXX2015}{XXX}

\title{On inferring structural connectivity from brain functional-MRI data}
%
%
%
%
%

\numberofauthors{3} 
%
\author{
%
%
\alignauthor
Somwrita Sarkar\\
       \affaddr{Design Lab}\\
       \affaddr{University of Sydney}\\
       \email{\normalsize somwrita.sarkar@sydney.edu.au}
\alignauthor
Sanjay Chawla\\
       \affaddr{Faculty of Engg and IT}\\
       \affaddr{University of Sydney}\\
       \email{\normalsize sanjay.chawla@sydney.edu.au}
\alignauthor 
Donna Xu\\
       \affaddr{Faculty of Engg and IT}\\
       \affaddr{University of Sydney}\\
       \email{\normalsize doxu2620@uni.sydney.edu.au}
}

\maketitle
\begin{abstract}
The anatomical structure of the brain can be observed via non-invasive techniques such as diffusion imaging. However, these are imperfect because they miss connections that are actually known to exist, especially long range inter-hemispheric ones. In this paper we formulate the inverse problem of inferring the structural connectivity of brain networks from experimentally observed functional connectivity via functional Magnetic Resonance Imaging (fMRI), by formulating it as a convex optimization problem. We show that structural connectivity can be modeled as an optimal sparse representation derived from the much denser functional connectivity in the human brain. Using only the functional connectivity data as input, we present (a) an optimization problem that models constraints based on known physiological observations, and (b) an ADMM algorithm for solving it. The algorithm not only recovers the known structural connectivity of the brain, but is also able to robustly predict the long range interhemispheric connections missed by DSI or DTI, including a very good match with experimentally observed quantitative distributions of the weights/strength of anatomical connections. We demonstrate results on both synthetic model data and a fine-scale 998 node cortical dataset, and discuss applications to other complex network domains where retrieving effective structure from functional signatures are important.    
%
%
\end{abstract}


\terms{Theory, Algorithms, fMRI, structural, functional, connectivity, networks}

\section{Introduction} 
\label{sec1}
Networks or graphs are used to model the structure and dynamics of many real world systems, such as the internet and the world wide web, scientific collaborations, infrastructure systems such as road, rail, power or transport networks, the brain, social networks of individuals, knowledge and design structures in engineering, information and knowledge dynamics in firms or organizations~\cite{albert2002, boccaletti2006, barthelemy2011, sporns2011, sarkar2013b, dong2015}. Usually, an implicit assumption is that the structure of the network is observable, and forward models or inference of dynamics can be based on the topological organization of the system, for example~\cite{buldyrev2010}. 

However, many of these systems, notably the brain, have structures that are difficult to map but can be observed via the activity they support. Further, knowledge of functional dynamics can also provide deeper information on active structure (as a subset of the total structure) and its temporal variation in systems. Therefore, we can define the \textit{inverse problem of deriving structure from the functional signature of dynamical networks}. The anatomical structure of brain networks can only be observed via techniques such as diffusion imaging, [e.g. Diffusion Spectrum Imaging (DSI), Diffusion Tensor Imaging (DTI)], which are imperfect because they miss edges that are known to actually exist in the system, especially long range ones between the two hemispheres~\cite{bullmore2009, sporns2011, robinson2012, robinson2014, deco2014}. Further, not all of the anatomical connections are active at all times; some are dormant, leading to the notion of \textit{effective connectivity}.  It is possible to observe functional connectivity in these systems, measured in terms of correlations of network activity via techniques such as functional Magnetic Resonance Imaging (fMRI), EEG, or MEG. Methods that can infer the anatomical, structural, or effective connectivities, starting from the functional signature will therefore provide a significant step forward in understanding the normal and/or diseased structural and functional states of the brain by providing insight into the links between structure and function. 

Solving either the forward problem (inferring dynamics from structure) or the inverse problem (inferring structure from dynamics) exactly is currently beyond the capacity of any discipline. However, recent efforts in the brain networks domain show that both the forward and inverse problems are extremely topical in several disciplines~\cite{robinson2012, robinson2014, honey2008, papalexakis2014, davidson2013} across computational neuroscience, data mining and physics. In this paper, we pose the inverse problem of mining the \textit{structural connectivity} of brain networks from a signature of its \textit{functional connectivity}. Different approaches in different disciplines model the problem in several ways. For example, computational neuroscience relies heavily on experiments followed by the use of established graph theoretic analysis~\cite{honey2008}, whereas physics based approaches rely on theoretical modeling of physiological phenomena, such as the expected behavior of eigenvalues of systems in stable states or near critical regimes~\cite{robinson2012, robinson2014, deco2014}. In data mining, recent approaches have focussed on deriving functional connectivity signatures from functional connectivity data, using sparse representations~\cite{papalexakis2014, davidson2013, sun2009}, but we are not aware of any work on deriving structural connectivity from functional connectivity using sparse representations. 

This paper addresses some significant gaps that remain unexplored in previous (very recent) attempts on modeling the inverse problem~\cite{robinson2014, deco2014}. Firstly, we focus on the sparse structure of structural connectivity as opposed to the dense structure of functional connectivity: an optimal solution will be sparse, since experimental measurements show that the fMRI signature is always much denser than the structural/anatomical network. Secondly, while it is reasonable to assume that functional connectivity ultimately derives from structure and so the two would share topological properties, we focus on the issue of the actual quantitative or numerical distributions of the connection strengths in both cases. As experimental data shows, these are very different for the structural and functional data, with the structural DSI or DTI data being non-negative and the fMRI data containing both positive and negative correlations. Additionally, the numerical ranges and the distribution of the data are very different for both cases.

In this paper, we propose a model for deriving a sparse weighted representation of the structural connectivity of the brain from its dense (full matrix) weighted functional connectivity. Our main contributions are as follows: 

\begin{enumerate}
\item \textbf{Formulation:} A convex optimization problem definition of deriving the structural connectivity of the brain from its fMRI connectivity.
\item \textbf{Algorithm:} An ADMM based algorithm for solving the optimization problem. 
\item \textbf{Application:} Applying the approach to real experimental DSI and fMRI data. 
\item \textbf{Prediction and validation:} To what extent does the inferred structural connectivity confer with the known structural connectivity in the cortex and can it provide any new information about anatomical connectivity? We compare our results with (a) the known short range anatomical connectivity of the cortex, and (b) the known (but missing from experimental data) long range inter-hemispheric connectivity of the cortex (prediction of missing links in the structural network). 
\end{enumerate}

\section{Background}
\label{sec2}

Before formalizing the problem, we present some observations that will form the motivation for our optimization formulation, and also establish the basis for why posing the problem within an optimization framework can have significant advantages:  

\begin{figure}
\centering
\epsfig{file=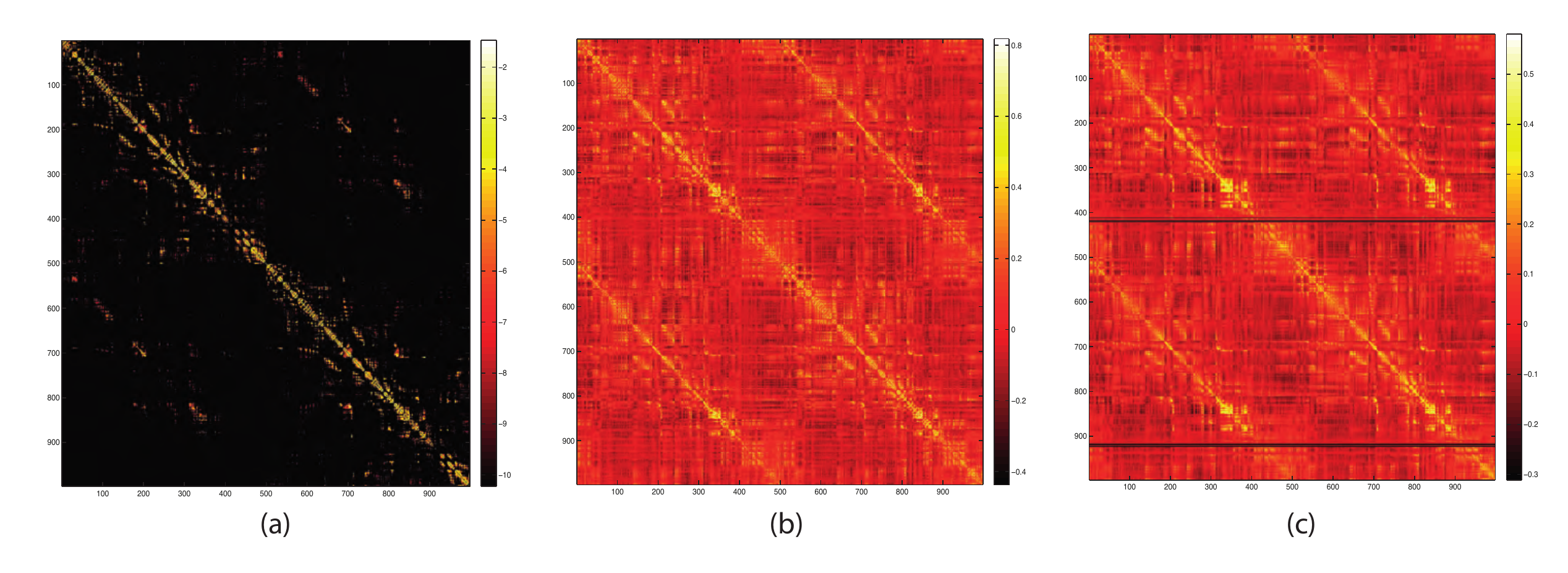, height = 1.2in, width = 3in}
\vspace{-0.8cm}
\caption{\label{Fig1} Experimentally measured anatomical and functional connectivity in the human cortex~\cite{hagmann2008}. (a) Structural connectivity measured via DSI. (b) Resting state functional connectivity measured via fMRI. (c) Linear correlation map between structural and functional connectivity matrices.}
\end{figure}

\begin{enumerate}
\item Experiments have shown that the structural and functional connectivity of the brain have high positive correlations with each other~\cite{honey2008}. In other words, if two areas of the brain are strongly connected anatomically, it is likely that they will share high functional correlations too. Figures~\ref{Fig1}(a) and (b) show the anatomical and resting state functional connectivity data from~\cite{hagmann2008}. Each row or column of the matrix shows a \textit{region of interest} or ROI of the cortex, and its entries represent structural or functional connectivity with all the other ROIs. The matrices represent both the left and right hemispheres, with the top half of the rows/columns representing one hemishphere and the bottom half the other. Figure~\ref{Fig1}(c) shows the linear correlation between the rows/columns of the structural and functional connectivity matrices.

\item Experiments show that the functional connectivity is based upon direct as well as indirect connections or the presence of both direct and indirect structural paths between brain regions~\cite{robinson2012, robinson2014, honey2008}. Thus, signatures of functional connectivity or correlations of activity as mapped via fMRI are significantly denser than signatures of structural connectivity as mapped via techniques such as DSI or DTI~\cite{hagmann2008}, since the functional connectivity is postulated to derive from both direct and indirect connectivity between brain regions. As can be seen from Figure~\ref{Fig1}(b), the functional signature is much denser than the structural one in Figure~\ref{Fig1}(a). Thus, it appears from experimental data that structural connectivity can be naturally expressed by a sparser representation while the functional one is a denser one. 
  
\item Current non-invasive techniques of mapping structural brain connectivity miss the long range connections, especially inter-hemispheric ones in the brain~\cite{robinson2012, robinson2014, honey2008, hagmann2008}. This is visible in Figs~\ref{Fig1}(a) and (b): while the main diagonal is similarly structured in both, the two off-diagonals, representing inter-hemispheric connectivity between the two hemispheres, are very weak in Fig~\ref{Fig1}(a) and strong in Fig~\ref{Fig1}(b), showing that the functional signature captures traces of long range connections via strong correlations that the structural data from DSI misses. 
\end{enumerate}

The above observations lead to the following informal characterization of the problem. Given a dense functional connectivity matrix, can we derive an optimal sparse representation of structural connectivity, by modeling physiologically and experimentally known constraints on structural and functional connectivity? Further, (a) how much of experimentally mapped and known structural connectivity is the sparse representation able to predict, and (b) can the sparse representation also predict structural connectivity that is known to exist anatomically, but is missing from data that comes from the current state of the art non-invasive imaging techniques?

\section{Problem Definition}
\label{sec3}  

Formally, we consider two graphs $G_{s} = (V,E_{s})$ and $G_{f} = (V, E_{f})$, with ROIs of the cortex represented as $N$ nodes in the set $V$, and the edge sets $E_{s}$ and $E_{f}$ representing structural and functional connectivity edges, respectively, between two ROIs or nodes. While structural connectivity is a measure of the actual anatomical connectivity between two ROIs or nodes, functional connectivity is a measure of correlation between activity on two nodes as the brain is in resting state or performs a specific task. Thus, note that while the presence of an edge in the set $E_{s}$ is a measure of actual physical connectivity, an edge in the $E_{f}$ signifies the level of correlation between activity based on the BOLD signature~\cite{sporns2011}. 

However, based on the observations in the background and previous research~\cite{meunier2009, meunier2010, sporns2011, bullmore2009, robinson2012, robinson2014, hagmann2008, honey2008, deco2014}, we assume that the structural and functional connectivity of a node to other nodes is positively correlated, as discussed in the Background section. More formally, we represent the experimentally measured structure and functional connectivity networks $G_{s}$ and $G_{f}$ by weighted adjacency matrices $S$ and $F$ respectively, where the entries $S_{ij}$ and $F_{ij}$ represent the structural and functional connection weights between brain regions $i$ and $j$, respectively. Thus, each column (or row) $s_{i}$ or $f_{i}$ represents the structural or functional connectivity of node $i$ to all the other nodes in the network. 

Our main task can be stated as inferring a sparse representation of $F$, as a representation of the structural connectivity, by modeling constraints that derive from known experimental and physiological observations. The simplest way of stating this is \vspace{-0.1cm} \begin{equation} FX = F, \end{equation} where $X$ represents our sparse representation matrix, and we model it as a transformation that takes F to itself. However, to extract the required pattern, this form can be further simplified to our advantage if we work with a lower dimensional representation of $F$ rather than the full $N \times N$ version. The reasoning for this is as follows. Row $i$ of $F$ shows the connectivity of node $i$ with all the other connections. Because of the physiologically known modularity structure of the brain (e.g. visual cortex, auditory cortex, etc.), we know that several other nodes will typically have connectivities very similar to node $i$. Thus, we are looking for a representation where each node can be represented as a point in $k-$dimensional space, with $k<<N$, such that if two nodes share similar connectivities to other nodes in $F$ or $S$, then they must lie very close to each other in the $k-$dimensional space. This can be achieved in several ways, but here we use the simplest possible representation: the spectral representation of $F$, or the co-ordinates given by the first $k$ eigenvectors of $F$; $F = VDV^{T}$, $F_{k} = V_{k}D_{k}V_{k}^{T}$, where $V_{k}$ which is $k \times N$ matrix
, now represents the positions of $N$ nodes in $k-$dimensional space $\mathbb{R}^{k}$. When the transformation $X$ is applied to $F$, the components of sparse $X$ will be like weights on the columns of $V_{k}$, ``picking out" the most relevant connections of node $i$ to all other nodes. Thus we have \vspace{-0.1cm} \begin{equation} V_{k}X = V_{k} \end{equation}. 

This is also the simplest possible representation for which the known physiological constraints described in the previous section can be captured: (a) high correlations in the fMRI data must imply high probability of structural connectivity, but (b) the structural connectivity must be sparse, since only direct connections must be inferred, and (c) both the short range and the long range structural connectivity must be inferred using the functional connectivity.

One complication is introduced by the presence of negative entries in the experimental fMRI data. The meaning of negative correlations in fMRI data is under open debate and has been attributed to (a) pre- and post-processing of the fMRI data (especially removals related to other physiological functions such as cardiovascular and respiratory functions from the raw fMRI data, and removals of global modes when temporal averaging is performed to reveal spatial correlations as demonstrated in Fig.~\ref{Fig1}(b), (b) BOLD activity and haemodynamic effects, and (c) excitatory versus inhibitory effects as correlated with positive or negative BOLD response. Since the exact meaning of the negative correlations in the fMRI matrix and its physiological significance is unclear and varied in the research literature, and all structural connectivity data as measured by DSI or DTI techniques is positive, we would like to separate out the positive and negative parts, as \vspace{-0.1cm} \begin{equation} V_{k}(X_{p} + X_{n}) = V_{k}, \end{equation} where $X_{p}$ is sparse and positive, and $X_{n}$ is small and negative. Separating out the positive and negative also makes our results easier to interpret. Surprisingly, we discover in our results that both the positive and negative solutions echo back a similar ``picture" of structural connectivity, but $X_{p}$ is much closer to predicting the real structural connectivity. The robustness of this result is particularly significant, because unlike previous efforts~\cite{deco2014}, we make no use of the structural connectivity matrix to predict the same in our optimization formulation; all our computations are done on the functional connectivity matrix and what we derive is then shown to be very close to the experimentally measured structural connectivity. 

Thus, the final optimization problem can be stated as \vspace{-0.2cm} \begin{equation} \begin{split} \mbox{min       } ||X_{p}||_{1} + \frac{\lambda_{n}}{2}||X_{n}||_{2}^{2} \\ \mbox{sub to       } V_{k}(X_{p} + X_{n}) = V_{k} \\ diag(X_{p}) = 0 \\ diag(X_{n}) = 0 \\ X_{p} \geq 0 \\ -X_{n} \geq 0. \end{split} \end{equation} Note that, similar to~\cite{elhamifar2013}, we want to enforce the diagonals of the solution variables to be 0, so as to avoid the trivial solution of each node expressing itself as its own linear combination and none of the others.

\section{Optimization Algorithm}
\label{sec4}

We present an ADMM based solution framework for solving the above problem. Introducing auxilliary variables $A$ and $B$ for the optimization variables $X_{p}$ and $X_{n}$, and indicator functions $I_{1+}(X_{p})$ and $I_{2+}(-X_{n})$ for the non-negative and negativity constraints, we have: \begin{equation} \begin{split}  L[A, B, X_{p}, X_{n}, \Delta_{1}, \Delta_{2}] = \\ ||X_{p}||_{1} + \frac{\lambda_{n}}{2}||X_{n}||_{2}^{2} + \frac{\lambda_{t}}{2} ||V_{k} - (A + B)V_{k}||_{2}^{2} \\ + \mbox{tr}[\Delta_{1}^{T} (A - (X_{p} - diag(X_{p}))] \\ + \mbox{tr}[\Delta_{2}^{T} (B - (X_{n} - diag(X_{n}))] \\ + \frac{\rho_{1}}{2} ||A - (X_{p} - diag(X_{p}))||_{2}^{2} \\ + \frac{\rho_{2}}{2} ||B - (X_{n} - diag(X_{n}))||_{2}^{2} \\ + I_{1+}(X_{p}) + I_{2+}(-X_{n}), \end{split} \end{equation} where $ I_{1+}(X_{p}) = 0$ when $X_{p} \geq 0$ and $\infty$ otherwise, and $ I_{2+}(-X_{n}) = 0$ when $X_{n} \leq 0$ and $\infty$ otherwise. 

Now we minimize $L$ (w.r.t. its arguments) using a standard iterative
ADMM process where we differentiate with respect to one variable while
keeping the others fixed and finally updating the Lagrange multipliers
$\Delta_{1}$ and $\Delta_{2}$.  We represent with $A^{*}, B^{*}, X_{p}^{*}, X_{n}^{*}, \Delta_{1}^{*}, \Delta_{2}^{*}$, the updated variables $A, B, X_{p}, X_{n}, \Delta_{1}, \Delta_{2}$.
The full algorithm is presented in Algorithm \ref{alg1}.


\subsection{Updating $A$}

$A$ can be updated as computing \begin{equation} \begin{split} \frac{\partial L}{\partial A} = \frac{\partial}{\partial A} \bigg[ \frac{\lambda_{t}}{2} || V_{k} - (A + B)V_{k}||^{2}_{2} \\ + \frac{\rho_{1}}{2} ||A - X_{p} + diag(X_{p})||^{2}_{2} \\ + tr \big[ \Delta_{1}^{T} (A - X_{p} + diag(X_{p}))\big] \bigg], \end{split} \end{equation} and setting it to 0. We then get \begin{equation} A^{*} = (\lambda_{t} V_{k}^{T}V_{k} + \rho_{1} I)^{-1} [\lambda_{t} V_{k}^{T} (V_{k} - BV_{k}) + \rho_{1} X_{p} - \Delta_{1} ]. \end{equation}

\subsection{Updating $B$}

The form for $B$ is exactly similar to $A$. Thus, \begin{equation} B^{*} = (\lambda_{t} V_{k}^{T}V_{k} + \rho_{2} I)^{-1} [\lambda_{t} V_{k}^{T} (V_{k} - AV_{k}) + \rho_{2} X_{n} - \Delta_{2} ]. \end{equation}    

\subsection{Updating $X_{p}$}

Since the $\ell_{1}-$norm of $X_{p}$ is to be minimized, along with the square terms, $X_{p}$ can be updated in closed form using a soft-thresholding operator as follows: \begin{equation} X_{p}^{*} = Y - diag(Y),  \end{equation} where \begin{equation} Y = T_{\frac{1}{\rho_{1}}} \bigg( A + \frac{\Delta_{1}}{\rho_{1}} \bigg), \end{equation} \begin{equation} = \bigg( A +\frac{\Delta_{1}}{\rho_{1}} - \frac{1}{\rho_{1}} 11^{T} \bigg)_{+} sgn \bigg(A + \frac{\Delta_{1}}{\rho_{1}} \bigg), \end{equation} with \begin{equation} T_{\eta}(v) = (|v| - \eta)_{+} sgn(v). \end{equation} Now applying the indicator function $I_{1+}(X_{p})$, \begin{equation} \begin{split} X_{p}^{*} = Y - diag(Y), \\ Y = \bigg( A +\frac{\Delta_{1}}{\rho_{1}} - \frac{1}{\rho_{1}} 11^{T} \bigg)_{+}, \end{split} \end{equation} where $Y$ takes only positive values due to $I_{1+}(X_{p})$. 

\subsection{Updating $X_{n}$}

Similar to the derivation for $A$ and $B$, we can differentiate with respect to $X_{n}$ as \begin{equation} \begin{split} \frac{\partial L}{\partial X_{n}} = \frac{\partial}{\partial X_{n}} \bigg[ \frac{\lambda_{n}}{2} ||X_{n}||_{2}^{2} \\ + \frac{\rho_{2}}{2} || B - X_{n} + diag(X_{n})||^{2}_{2} \\ + tr \big[ \Delta_{2}^{T} (B - X_{n} + diag(X_{n})) \big] \bigg], \end{split} \end{equation} set the derivative to 0 and apply indicator function $I_{2+}(-X_{n})$ to get \begin{equation} X_{n}^{*} = \bigg( \big[ \frac{1}{\lambda_{n} + \rho_{2}} I \big] \big[\rho_{2}B + \Delta_{2} \big] \bigg)_{-}, \end{equation} where $I$ is the identity matrix and $X_{n}$ takes on only negative values because of $I_{2+}(-X_{n})$.  

\subsection{Updating $\Delta_{1}$ and $\Delta_{2}$}

Finally, we update the Lagrange multipliers as \begin{equation} \Delta_{1}^{*} = \Delta_{1} + \rho_{1} ( A^{*} - X_{p}^{*}), \end{equation} and \begin{equation} \Delta_{2}^{*} = \Delta_{2} + \rho_{2} ( B^{*} - X_{n}^{*}). \end{equation}

\begin{algorithm}
\begin{algorithmic}[1]
\REQUIRE
$F, k$,$\lambda_{t},\rho_{1},\rho_{2}$ \\
\ENSURE
$X_{p},X_{n}$ \\
Initialize $X_{p},X_{n},A, B, \Delta_{1},\Delta_{2}$ \\
\STATE $F \leftarrow  VDV^{T}$
\STATE Use $V_{k}$ from $V_{k}D_{k}V^{T}_{k}$ (first $k$ eigenvectors)
\REPEAT
\STATE $A \leftarrow 
(\lambda_{t} V_{k}^{T}V_{k} + \rho_{1} I)^{-1} [\lambda_{t} V_{k}^{T} (V_{k} - BV_{k}) + \rho_{1} X_{p} - \Delta_{1} ]$ \\
\STATE $B \leftarrow
(\lambda_{t} V_{k}^{T}V_{k} + \rho_{2} I)^{-1} [\lambda_{t} V_{k}^{T} (V_{k} - AV_{k}) + \rho_{2} X_{n} - \Delta_{2} ] $ \\
\STATE $Y \leftarrow  \bigg( A +\frac{\Delta_{1}}{\rho_{1}} - \frac{1}{\rho_{1}} 11^{T} \bigg)_{+} $\\
\STATE $X_{p} \leftarrow Y - diag(Y)$ \\
\STATE $Z \leftarrow
\bigg( \big[ \frac{1}{\lambda_{n} + \rho_{2}} I \big] \big[\rho_{2}B + \Delta_{2} \big] \bigg)_{-} $ \\
\STATE $X_{n} \leftarrow Z - diag(Z)$
\STATE
$\Delta_{1} \leftarrow 
\Delta_{1} + \rho_{1} ( A^{*} - X_{p}^{*})$ \\
\STATE
$\Delta_{2} \leftarrow
\Delta_{2}^{*} = \Delta_{2} + \rho_{2} ( B^{*} - X_{n}^{*})$

\UNTIL{convergence}
\STATE $X_{n} \leftarrow \frac{X_{n} + X_{n}^{T}}{2}$ \\
\STATE $X_{p} \leftarrow \frac{X_{p} + X_{p}^{T}}{2}$

\end{algorithmic}
\caption{ADMM Algorithm}
\label{alg1}
\end{algorithm}

\section{Results}

\subsection{Synthetic baseline model - I}

\begin{figure}
\centering
\epsfig{file=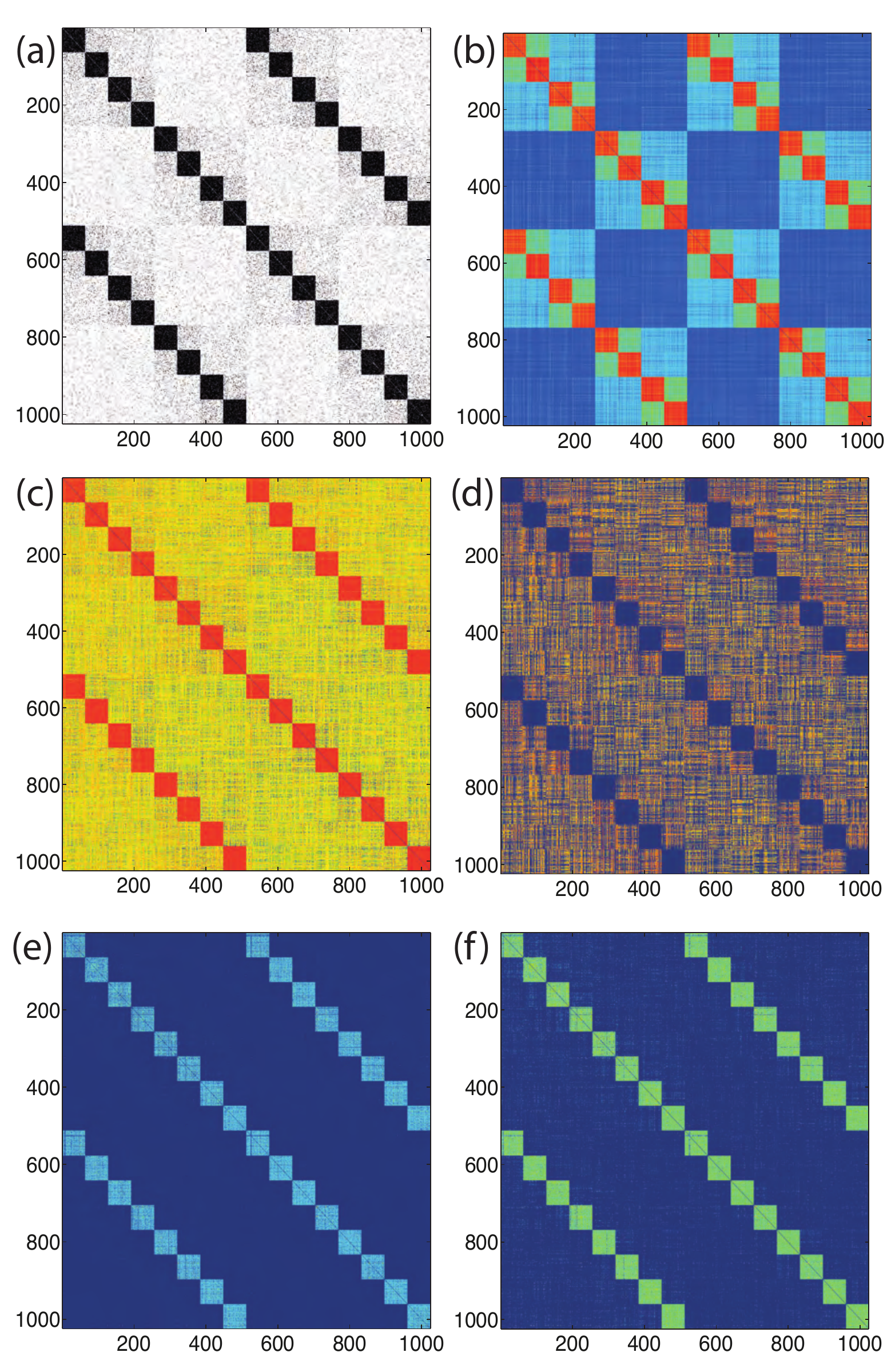, height = 4.2in, width = 3in}
\vspace{-0.5cm}
\caption{\label{Fig2} Synthetic baseline model I. (a) adjacency matrix for a hierarchical network with brain like architecture. (b) derived synthetic functional matrix. (c) $X_{p}$ (d) $X_{n}$ (e) $X_{pt}$ with near zero entries removed. (f) $X_{pn} = X_{p} \cap (X_{n} = 0)$. Parts (c), (d), (e), and (f) all bring out the connectivity structure of the adjacency matrix from the functional matrix. Although the hierarchical structure is not particularly visible in (e) and (f), it is detected, as confirmed by precision and recall measurements.}
\end{figure}

\begin{figure}
\centering
\epsfig{file=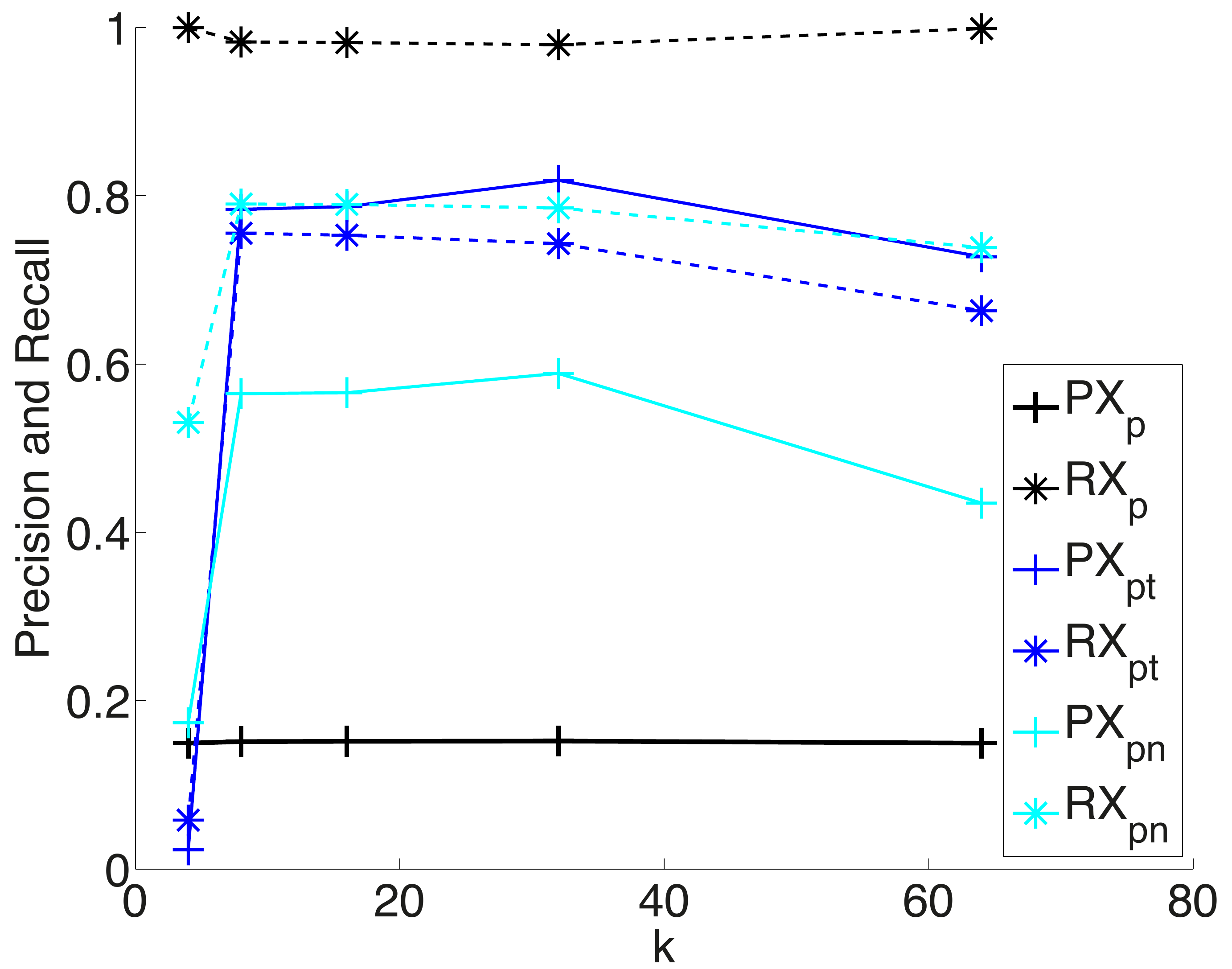, height = 2in, width = 2.8in}
\vspace{-0.5cm}
\caption{\label{Fig3} Precision and recall with respect to varying number of dimensions preserved in spectral representation $k$ for synthetic baseline model I. Solid lines show precision, dotted lines show recall, for $X_{p}$, $X_{pt}$ with near zero entries removed and $X_{pn} = X_{p} \cap (X_{n}=0)$. $X_{p}$ has low precision and almost perfect recall. Precision climbs significantly for $X_{pt}$ and $X_{pn}$ for lower recall of upto about 0.80.} 
\end{figure}

As discussed in the Background section, the real brain data that we will work with has known missing entries, especially for long range interhemispheric connections. In order to test and validate the approach on a baseline, we have generated synthetic models mimicing the known large scale architecture of the cortex. Studies have shown that the cortex has a hierarchical modular architecture with larger modules nesting smaller modules on several levels~\cite{robinson2009, meunier2009, meunier2010}. We have therefore generated hierarchically modular networks with stochastic block model type architectures, following closely the models introduced in~\cite{sarkar2013a, robinson2009}. This is the structural connectivity matrix $S$. Figure~\ref{Fig2}(a) shows an example hierarchical network adjacency matrix $S$, with 1024 nodes, and 16 modules at the finest scale along the main diagonal. Other hierarchical levels are formed by progressively reducing the probability of connectivity for 8 modules, 4 modules, and finally 2 modules representing the left and right hemishpheres. Further, a similar architecture is repeated for inter-hemispheric connectivities, appearing as the off-diagonals in the matrix. 

In order to generate a similar synthetic functional activity / correlations network, we have used the idea already observed in the Background section and Fig.~\ref{Fig1}: that the existence of direct as well as indirect paths in the structural matrix are highly positively correlated with measured functional activity in the cortex. That is, if two brain regions are connected by direct or indirect paths, then it is highly likely that their functional activity correlation signature will also be high. Shorter the path, more direct the connection, higher the probability for positive functional correlations. Following previous work done in~\cite{robinson2014, robinson2012} and experimental observations in~\cite{honey2008, deco2014}, a simple model of producing a synthetic functional matrix from a given structural matrix is defined as:\begin{equation} \label{eq_series} F = S + S^{2} + ... + S^{j}, \end{equation} where each power matrix captures the numbers of paths of length $j$ in the adjacency matrix. Analytically $j$ can go upto infinity, but practically and computationally, a small number can be chosen to simulate this. The reasoning behind this is twofold: (a) it can be assumed that higher powers (longer and longer paths) contribute less and less to the functional activity, and (b) the brain operates at marginal stability, close to but lower than the critical boundary between stable and unstable, thus implying the condition that the principal eigenvalue of $S$ will always be less than 1. In such a case, the series in Eqn~[\ref{eq_series}] converges, as shown in~\cite{robinson2014}. For more analytical detail, please refer to previous work in~\cite{robinson2012, robinson2014}. 

Figure~\ref{Fig2}(b) shows a synthetic functional connectivity matrix derived from the structural connectivity matrix by summing upto paths of length $5$. This is matrix of functional connectivity, denoted by $F$. 

We apply our algorithm onto the spectral representation of $F$, and perform an analysis on the structural connectivity extracted when we vary $k$, the number of dimensions preserved in the spectral representation. Figures~\ref{Fig2}(c) and (d) show the matrices $X_{p}$ and $X_{n}$, respectively at $k=16$. Further, $X_{p}$ shows a sparse structure with several entries extremely close to zero (by at least two or more orders of magnitude). We remove these near zero values, producing the matrix $X_{pt}$, Fig.~\ref{Fig2}(e), that brings out the structure of the adjacency matrix. Further, as is seen in Figs~\ref{Fig2}(c) and (d), both $X_{p}$ and $X_{n}$ bring out the structure in two different ways: the positive entries in $X_{p}$ bring out structure, and the zero and near-zero negative entries of $X_{n}$ also mimic and reinforce the same structure, since they imply the absence of negative relationships between the nodes. In the case of the synthetic baseline we work with here, the negative entries of $X_{n}$ are quite small, since there are no negative entries in $F$. But as we will see in the next section, $X_{n}$ serves an important role when $F$ contains negative entries. Thus, we can combine information by extracting $X_{pn} = X_{pt} \cap (X_{n} = 0)$, which is shown in Fig.~\ref{Fig2}(f). All these interpretations bring out similar architectures for the adjacency matrix $S$, starting from the functional matrix $F$. Observe that we make no use of $S$ to derive $X_{p}$, $X_{n}$, $X_{pt}$, and $X_{pn}$. 

Figure~\ref{Fig3}(a) shows the precision and recall for $X_{p}$, $X_{pt}$, and $X_{pn}$. Precision is the number of correctly identified connections in our solution divided by the total number of identified connections. Recall is the number of correctly identified connections divided by the total actual number of connections that actually exist in the structural connectivity matrix and should be ideally identified. $X_{p}$ has low precision and almost perfect recall. Precision climbs significantly for $X_{pt}$ and $X_{pn}$ for lower recall of upto about 0.80. This observation will become important in the next section, because for real brain data, precision will drop while recall will remain high. The reason for this is that we will predict several missing links with the algorithm, notably the inter-hemispheric connectivity, which will necessarily bring down the precision. Therefore, it was important to check with a baseline model that had all the relevant connectivity been present in the structural data, both precision and recall would have had acceptably high values.

\subsection{Brain data and data-driven baseline - II}

\begin{figure}
\centering
\epsfig{file=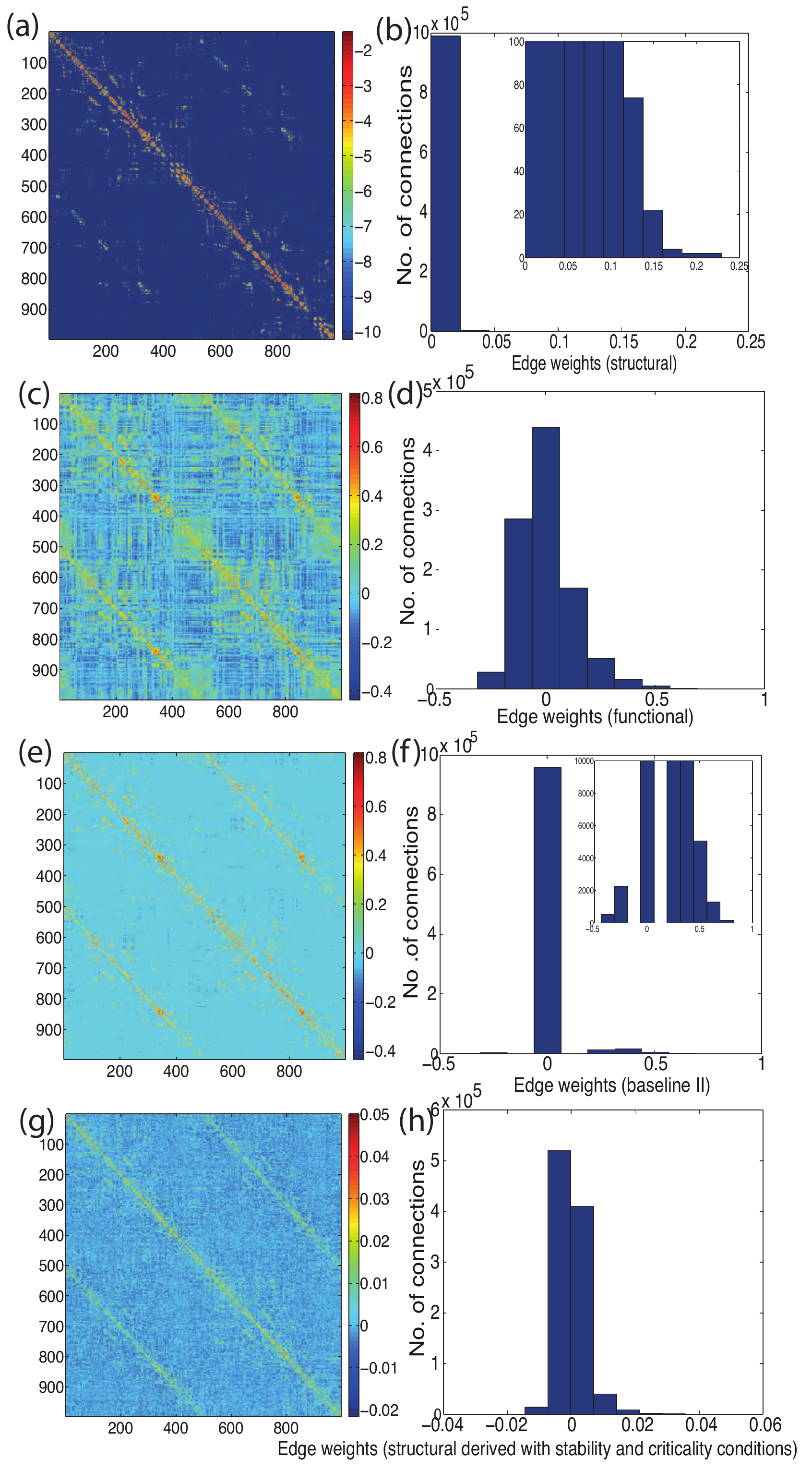, height = 5.9in, width = 3.4in}
\vspace{-0.5cm}
\caption{\label{Fig4} Brain data from~\cite{hagmann2008} and comparisons with baseline solution and solution from~\cite{robinson2014}. (a) Structural connection matrix. (b) Histogram for (a). (c) fMRI functional connection matrix. (d) Histogram for (b). (e) Simple thresholding based baseline solution derived from (c). (f) Histogram for (e) showing that simple thresholding may bring out topological connections weakly, but fails on connection strengths. (g) solution from~\cite{robinson2014}. (h) histogram for (g): solution in~\cite{robinson2014} correctly retrieves interhemispheric connectivity with largest eigenvalue of derived structural connectivity $=$ 0.87 that respects stability criteria, but is not sparse, has negative entries, and has a different range of connection strengths as compared to (b).}
\end{figure}

The brain data that we work with to test our algorithm comes from~\cite{hagmann2008}. A primary reason for choosing this dataset is that it is the only dataset we could find that maps structural (DSI) and functional (fMRI) connectivity for a defined set of 998 region parcellation map of the human cortex. While there are other datasets available that contain either only structural connectivity or only functional connectivity, it is often hard to put together these data and transfer results of one dataset to another since the parcellations used may be different in every case. These problems are discussed in~\cite{kong2013}. A second reason is that our results and methods could be tested against~\cite{deco2014, robinson2014} that also use the same dataset for the same aim as ours. Thirdly, while there is work on task based or disease based functional networks with temporal slices of functional connectivity, for e.g.~\cite{sun2009, papalexakis2014, davidson2013}, the problem that we were working on involves the use of temporally averaged spatial correlations between brain regions, such that the structural/physical/spatial connectivity can be derived. For this reason, we have used temporally averaged, resting state fMRI data, along with a parallel map of DSI based structural connectivity data for the same set of 998 regions of the cortex. Figure~\ref{Fig4}(a) and (c) show the structural DSI and functional fMRI data, respectively, with the histograms of their connections shown in Figs~\ref{Fig4}(b) and (d). Howeve, due to the general nature of the way we have formulated the problem, our algorithm could easily be applied to temporal data networks or disease networks, with the expectation that the effective structural connectivity or abnormal structural connectivity, respectively, could possibly be retrived from functional signatures of task based fMRI data or diseased brain fMRI data.   

The first baseline question we now ask is: if structure and functional data show high positive correlation, as claimed by the literature and discussed in the Background section, and is somewhat confirmed by a superficial visual look at the two datasets, can we apply a much simpler thresholding algorithm to derive the structural connectivity from the functional one? We tested this using Algorithm 2. 

\begin{algorithm}
\begin{algorithmic}[1]
\REQUIRE
$F$, $S$ \\
\ENSURE
$S*$ \\
\STATE $z \leftarrow$ Compute the number of non-zero entries in $S$
\STATE $o \leftarrow$ Order entries of F (+ve and -ve) by magnitude in descending order
\STATE $t \leftarrow o(z)$
\STATE $S* \leftarrow$ (|F|>t) 
\STATE Compare properties of S and S*
\end{algorithmic}
\caption{Simple Thresholding Algorithm}
\end{algorithm}

The results are shown in Fig.~\ref{Fig4}(e) and (f). Note that we cannot compute precision and recall in any reliable way for the derived $S*$, because we have used the number of non-zero entries in $S$ to decide the threshold $t$. The results show, however, that even though the visual form of $S*$ in Fig.~\ref{Fig4}(e) does have some similarity to $S$ in Fig.~\ref{Fig4}(a), the histograms of connection strengths look quite different. This happens because thresholding by absolute value to bring the sparsity of $S*$ close to the sparsity of $S$ causes values to intermittently disappear from $F$, since $F$ contains both negative and positive values. Further, this simple thresholding mechanism uses information in $S$, which our algorithm does not.

Finally, we also study results from~\cite{deco2014} and~\cite{robinson2014}. The results in~\cite{deco2014} have used the coarser $66 \times 66$ region definition, and are therefore not as fine grained as the $998 \times 998$ region definitions used in our work (the same source data is available at both resolutions, but using the higher resolution is more challenging because it introduces much more fine structure and constraints into the problem). The results from~\cite{robinson2014} are shown in Fig.~\ref{Fig4}(g). The biggest strength of this result is that there is a physiologically based model that explains both the forward and inverse problems, along with predicting the missing inter-hemispheric connectivity, the predicted structural connectivity matrix had its largest eigenvalue at 0.87, which is in experimental agreements with EEG recordings of brain activity, and satisfies the theoretical and physiological conditions of dynamical stability and criticality in normal brain functioning. However, this solution was not sparse and returns a full matrix, has negative entries, and shows a range of connection strengths different from the range in $S$, Fig.~\ref{Fig4}(h).     

\subsection{Known structure recovery and prediction of inter-hemispheric links}

\begin{figure}
\centering
\epsfig{file=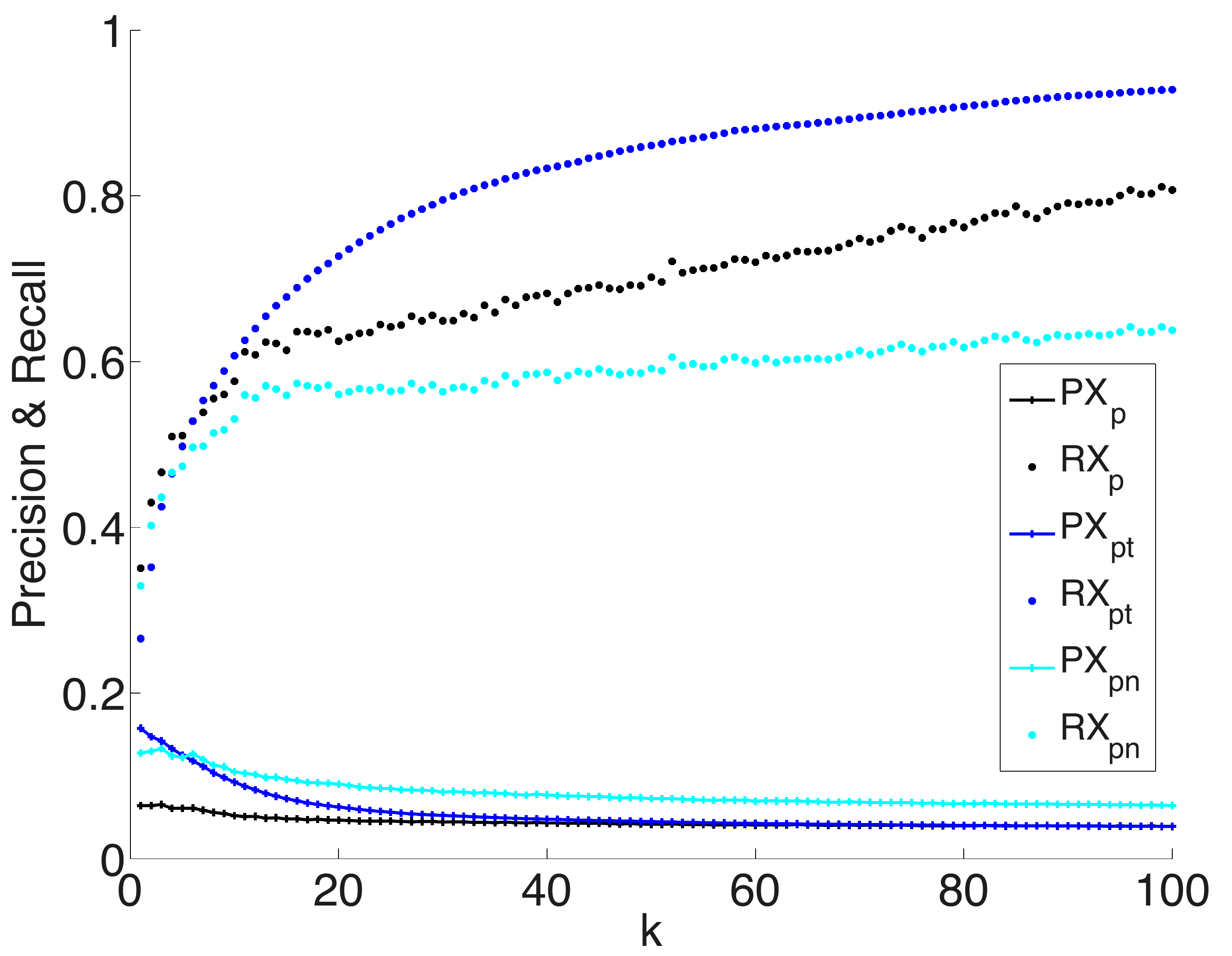, height = 2.4in, width = 3in}
\vspace{-0.5cm}
\caption{\label{Fig5} Precision and recall with respect to varying number of dimensions preserved in spectral representation $k$ for real brain data from~\cite{hagmann2008}. Solid lines show precision, dotted lines show recall, for $X_{p}$, $X_{pt}$ with near zero entries removed and $X_{pn} = X_{p} \cap (X_{n}=0)$. All have low precision, since original data has missing inter-hemispheric connections and solutions identify this strongly. Recall is high for all solutions, with $X_{pt}$ showing best performance.}
\end{figure}

\begin{figure*}
\centering
\epsfig{file=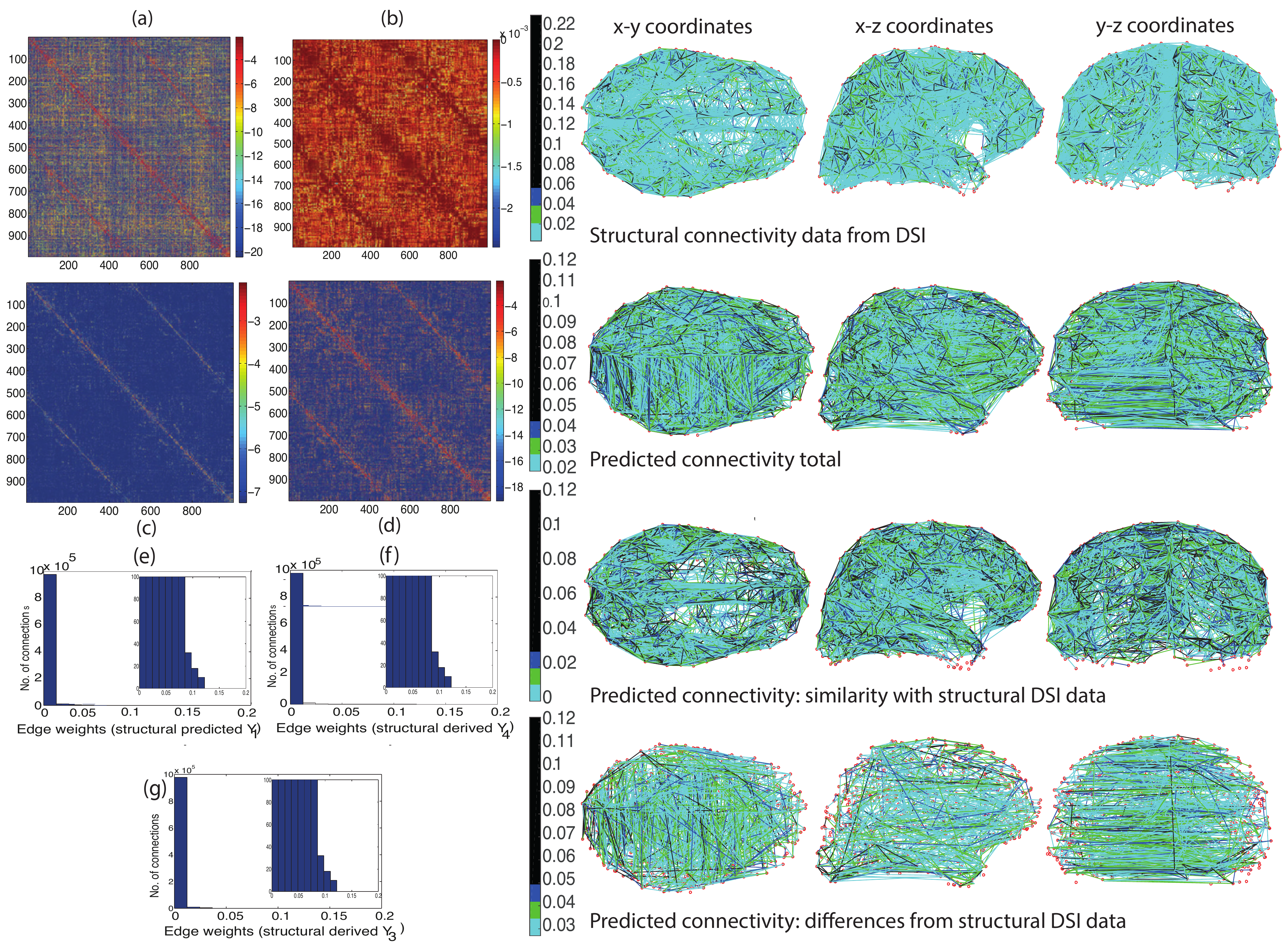, height = 5.1in, width = 7in}
\vspace{-0.5cm}
\caption{\label{Fig6} Results of applying algorithm on real brain data. (a) $X_{p}$. (b) $X_{n}$. (c) $X_{pt}$. (d) $X_{pn}$. (e), (f), (g) show histograms of connections strengths for (a), (c), and (d), respectively. Panel on the side shows comparison of actual and predicted connectivity. Columnwise: the plots represent 2D plots of nodes and edges with x-y, x-z, and y-z coordinates showing three different visualizations of the cortex. Row-wise: Top row shows the edges from the DSI structural connectivity matrix $S$, second row shows the edges from the solution matrix $X_{pt}$ shown in Fig.~\ref{Fig6}(c); note that the solution shows long range connections. Third row shows the edges from $X_{pt} \cap S$, the part of $S$ that is detected in $X_{pt}$. Last row shows the links that are predicted in $X_{pt}$ but do not exist in $S$, note the long range connections. For ease of representation, solution plots have been thresholded at 0.02, so that too many very low weight edges are not visible. Our retrieved solutions have more connections predicted for both intra-hemispheric and inter-hemispheric connections.}
\end{figure*}


We now present the results of our algorithm applied to the brain data from~\cite{hagmann2008}. Figures~\ref{Fig6}[a-d] show the inferred matrices $X_{p}$, $X_{n}$, $X_{pt}$, and $X_{pn}$, respectively. From visual observation, we see that both intra-hemispheric (strong main diagonal connectivity) and inter-hemispheric connectivity (strong off-diagonals) between the left and right hemispheres has been retrived. Figures~\ref{Fig6}[e-f] show the histogram of connection strengths for $X_{p}$, $X_{pt}$, and $X_{pn}$, respectively. The range of connection strengths inferred is very close to the distribution in the experimental DSI based structural connectivity matrix, Fig~\ref{Fig4}(a) and (b). Our solutions predict a higher number of both intra-hemispheric and inter-hemispheric connections as compared to the existing structural matrix. The panel of plots on the right in Fig.~\ref{Fig6} shows the actual and predicted structural connectivities, along with plots of similarities and differences between actual and predicted connections. The similarity plot establishes that a high number of actually existing connections are recovered, as also confirmed by recall computation, and the differences plot establishes that inter-hemispheric long range connections are retrieved. Note again, that we make no use of the experimental structural connectivity matrix to make our inferences, using only the functional connectivity data and applying our algorithm to it. 

Further, we also compare the results for precision and recall in comparison with the experimental structural connectivity matrix. For all three matrices, $X_{p}$, $X_{pt}$, and $X_{pn}$, recall is high, with $X_{pt}$ performing the best at about 0.7 - 0.8. It would appear that increasing the value of $k$ further results in even better or near perfect recall. However, we keep in mind the related important task of predicting the inter-hemispheric connectivity, that does not have a strong signature in the experimental structural connectivity data. Therefore, increasing the value of $k$ results in the solutions moving closer and closer to the experimental structural matrix with decreasing intensities of entries for the inter-hemispheric connectivity. This observation also explains why precision is low for all three matrices $X_{p}$, $X_{pt}$, and $X_{pn}$: precision is computed as the ratio of the correctly detected \textit{existing and mapped} connections in the experimental structural DSI data to the total number detected. Because the predicted interhemispheric connections retreived are comparable in both number and intensity to the intra-hemispheric ones, but do not exist in the existing and mapped structural DSI data, the precision goes down. However, we have shown in the previous section on the baseline synthetic model, that when no connections are missing, precision and recall return sufficiently high acceptable values.

\section{Related Work}
Brain network analysis is an active area of research across
many disciplines including data mining and machine learning,
complex systems and of course neuroscience~\cite{bullmore2009, sporns2011, deco2014, robinson2014, kong2013, papalexakis2014, davidson2013, sun2009}. 
However, an
optimization based perspective has been mostly pursued
in the data mining and machine learning community~\cite{sun2009,davidson2013,papalexakis2014}. These have mostly focused on reconstructions and inferences of functional connectivity from various types of functional connectivity data. 

A popular approach has been to represent
neuroimaging or functional data as tensors. For example, fMRI data can be
represented as a fourth order tensor consisting of the 3D spatial coordinates
and time. Data mining tasks, like
supervised learning, have been extended for tensors for brain data analysis. 
In ~\cite{he2014},
data is assumed to be given in the form $(X_i,y_{i})_{i=1}^{n}$,
where $X_{i}$ is the tensor representing an fMRI and $y_{i}$ 
is a binary label indicating whether the brain sample is normal
or suffering from a disease (like ADHD). Then an SVM optimization
formulation is extended to learn the weight tensor of the separating
hyperplane. 

In Davidson et. al.~\cite{davidson2013} a tensor formulation is used to
set up an optimization problem for both node and edge discovery.
The motivation is that fMRI parcellation needs to be aggregated
at the appropriate level which simultaneously corresponds to
coherent functional regions while keeping the fMRI activity
discriminative enough to capture local variations.

The body of work introduced in ~\cite{huang2011,sun2009} uses
an optimization formulation to estimate a sparse
inverse covariance matrix from the sample covariance matrix
extracted from fMRI data.
The insight for infering the inverse covariance matrix is
the well known result that if data follows a multivariate normal distribution
then the zero locations in the inverse covariance matrix corresponds
to conditional independence between the variables, i.e., if $\Theta(i,j) = 0$
then $i$ and $j$ are conditionally independent and thus there
does not exist an  edge between them.  Sparsity is enforced using the
$\ell_{1}$ regularization on $vec(\Theta)$.  However, the focus is still on the retrieval of relevant functional connectivity from data on functional connectivity. 

The inverse problem of inferring structural or anatomical connectivity is extremely topical in the computational neuroscience community~\cite{deco2014, robinson2012, robinson2014}, but to the best of our knowledge, has not been modeled with an optimization framework in the machine learning and data mining community. One of our chief contributions is that we show that there can be significant improvements in the quality of the predictions if we model the inverse problem using optimization. Our optimization formulation is also quite distinct in that we assume that the spectral representation
of the fMRI adjacency matrix lives in a union of low-dimensional spaces where
data points can be expressed as a sparse linear combination of 
other elements in the subspace. This has been referred to as the self-expressive property of data in ~\cite{elhamifar2013}, and we show that it is possible to derive an optimal sparse representation of structural/anatomical connectivity, starting from functional connectivity data using this property.

\section{Conclusions}
\label{seclast}

We presented a convex optimization formulation and ADMM algorithm for solving the inverse problem of deriving the sparse structural connectivity of the brain from its functional connectivity signature. To the best of our knowledge, this inverse structural/anatomical inference problem from fMRI data  has not been modeled within an optimization framework before.

Using only the functional connectivity data as input, we presented (a) an optimization problem that models constraints based on known physiological observations, and (b) an ADMM algorithm for solving it. We showed that the algorithm not only successfully recovers the known structural connectivity of the brain, but is also able to robustly predict the long range interhemispheric connections missed by DSI or DTI. Particularly, it addresses two principal gaps that remained unexplored in previous attempts: (a) our solution was sparse, and was very close to the sparsity that is actually observed in a DSI or DTI based experimental anatomical network, and (b) our predictions of the numerical distributions of the weights or connections strengths also showed a good match with the DSI or DTI based experimental anatomical network, even though we use only the fMRI data that has a very different distribution of weights. Further, we demonstrated these results on one of few available datasets that contain parallel maps of structural and functional connectivity of the human cortex parcellated into fine-scale 998 regions. Previous results have either been shown on the much coarser 66 node parcellation, and/or has not had a sparse form with close matches to the actual numerical distributions of the experimentally observed connection weights. 

For future work, it will be possible to extend this formulation to include constraints coming in from the structural data and constraints that also respect the criticality and stability conditions that have been noted to be important in both~\cite{deco2014} and~\cite{robinson2014}. For example, the problem could be formulated as a constrained matrix completion problem, where we write an optimization problem to use the fMRI data to introduce links into the DSI data in a constrained way, while also forcing the stability and criticality conditions on eigenvalues as well as maintaining the required amount of sparsity and the quantitative distribution of weights/connection strengths. The method could also be extended to model more realistic cases of asymmetric connectivity and/or combinations of excitatory and inhibitory connectivity in the cortex, since in each of these cases, the basic optimization formulation could be changed to model different types of conditions.       

While the present paper focusses on brain networks, the inverse problem of deriving structure from function in general is relevant to many other types of dynamical networks. Infrastructure networks, such as roads or communication lines, have links that can support fixed flows. In this case, the structure of carrying capacity can be mapped exactly, but functional data on congestion and jams can be used to infer management or design strategies for the future, for altering the physical structure of the network, or altering the flow or capacities of links~\cite{pang2013}. As a second example, consider that while the exact structure of online social networks such as Facebook can be mapped, much of the structure may be dormant, as not all links are active all of the time. In such a case, deriving the \textit{effective structure} of the network via studying time-stamped dynamics such as the frequency of interactions via specific links, can provide much deeper levels of information about the structure of the network. Thus, we believe that modeling this problem within an optimization framework as a general strategy can provide significant advantages and deep insights for several domains.

\section{Acknowledgements}
The authors thank Prof. P.A. Robinson and Dr K. Aquino for valuable comments and insights. 
\bibliographystyle{abbrv}
\bibliography{S}  

\begin{thebibliography}{10}

\bibitem{albert2002}
R.~Albert and A.~L. Barabasi.
\newblock Statistical mechanics of complex networks.
\newblock {\em Rev. Mod. Phys.}, 74:47--97, 2002.

\bibitem{barthelemy2011}
M.~Barthelemy.
\newblock Spatial networks.
\newblock {\em Phys. Rep.}, 499:1--101, 2011.

\bibitem{boccaletti2006}
S.~Boccaletti, V.~Latora, Y.~Moreno, M.~Chavez, and D.~U. Hwang.
\newblock Complex networks: Structure and dynamics.
\newblock {\em Physics Reports}, 424:175--308, 2006.

\bibitem{buldyrev2010}
S.~V. Buldyrev, R.~Parshani, G.~Paul, H.~E. Stanley, and S.~Havlin.
\newblock Catastrophic cascades of failures in interdependent networks.
\newblock {\em Nature}, 464:10.1038/nature08932, 2010.

\bibitem{bullmore2009}
E.~T. Bullmore and O.~Sporns.
\newblock Complex brain networks: graph theoretical analysis of structural and
  functional systems.
\newblock {\em Nat. Rev. Neurosci.}, 10:186--198, 2009.

\bibitem{davidson2013}
I.~Davidson, S.~Gilpin, O.~Carmichael, and P.~Walker.
\newblock Network discovery via constrained tensor analysis of fmri data.
\newblock In {\em ACM SIGKDD}, pages 194--202. ACM, 2013.

\bibitem{deco2014}
G.~Deco, A.~R. McIntosh, K.~Shen, R.~M. Hutchinson, R.~S. Menon, S.~Everling,
  P.~Hagmann, and V.~K. Jirsa.
\newblock Identification of optimal structural connectivity using functional
  connectivity and neural modeling.
\newblock {\em The Journal of Neuroscience}, 34(23):7910--7916, 2014.

\bibitem{dong2015}
A.~Dong and S.~Sarkar.
\newblock Forecasting technological progress potential based on the complexity
  of product knowledge.
\newblock {\em Technological Forecasting and Social Change}, 90B:599--610,
  2015.

\bibitem{elhamifar2013}
E.~Elhamifar and R.~Vidal.
\newblock Sparse subspace clustering: Algorithms, theory and applications.
\newblock {\em IEEE Transactions on Pattern Analysis and Machine Intelligence},
  35(11):2765--2781, 2013.

\bibitem{hagmann2008}
P.~Hagmann, L.~Cammoun, X.~Gigandet, R.~Meuli, C.~J. Honey, and V.~J. Weeden.
\newblock Mapping the structural core of the human cerebral cortex.
\newblock {\em PLOS Biol.}, 6(7):e159, 2008.

\bibitem{he2014}
L.~He, X.~Kong, P.~S. Yu, X.~Yang, A.~B. Ragin, and Z.~Hao.
\newblock Dusk: {A} dual structure-preserving kernel for supervised tensor
  learning with applications to neuroimages.
\newblock In {\em Proceedings of the 2014 {SIAM} International Conference on
  Data Mining, Philadelphia, Pennsylvania, USA, April 24-26, 2014}, pages
  127--135, 2014.

\bibitem{honey2008}
C.~J. Honey, O.~Sporns, L.~Cammoun, X.~Gigandet, J.~P. Thiran, R.~Meuli, and
  P.~Hagmann.
\newblock Predicting human resting-state functional connectivity from
  structural connectivity.
\newblock {\em PNAS}, 106(6):2035--2040, 2008.

\bibitem{huang2011}
S.~Huang, J.~Li, J.~Ye, T.~Wu, K.~Chen, A.~Fleisher, and E.~Reiman.
\newblock Identifying alzheimer's disease-related brain regions from
  multi-modality neuroimaging data using sparse composite linear discrimination
  analysis.
\newblock In {\em Advances in Neural Information Processing Systems 24: 25th
  Annual Conference on Neural Information Processing Systems 2011. Proceedings
  of a meeting held 12-14 December 2011, Granada, Spain.}, pages 1431--1439,
  2011.

\bibitem{kong2013}
X.~Kong and P.~S. Yu.
\newblock Brain network analysis: a data mining perspective.
\newblock {\em ACM SIGKDD Explorations Newsletter}, 15(2):30--38, 2013.

\bibitem{meunier2010}
D.~Meunier, R.~Lamboitte, and E.~T. Bullmore.
\newblock Modular and hierarchically modular organization of brain networks.
\newblock {\em Front. Neurosci.}, 4:Article 200, 2010.

\bibitem{meunier2009}
D.~Meunier, R.~Lamboitte, A.~Fornito, K.~Ersche, and E.~T. Bullmore.
\newblock Hierarchical modularity in human brain functional networks.
\newblock {\em Front. Neuroinform.}, 3:Article 37, 2009.

\bibitem{pang2013}
L.~X. Pang, S.~Chawla, W.~Liu, and Y.~Zheng.
\newblock On detection of emerging anomalous traffic patterns using gps data.
\newblock {\em Data and Knowledge Engineering}, 87:357--373, 2013.

\bibitem{papalexakis2014}
E.~E. Papalexakis, A.~Fyshe, N.~D. Sidiropoulos, P.~P. Talukdar, T.~M.
  Mitchell, and C.~Faloutsos.
\newblock Good-enough brain model: Challenges, algorithms and discoveries in
  multi-subject experiments.
\newblock In {\em ACM SIGKDD}. ACM, 2014.

\bibitem{robinson2012}
P.~A. Robinson.
\newblock Interrelating anatomical, effective, and functional brain
  connectivity, using propagators and neural field theory.
\newblock {\em Phys. Rev. E}, 85:011912, 2012.

\bibitem{robinson2009}
P.~A. Robinson, J.~A. Henderson, E.~Matar, P.~Riley, and R.~T. Gray.
\newblock Dynamical reconnection and stability constraints on cortical network
  architecture.
\newblock {\em Phys. Rev. Lett.}, 103:108104, 2009.

\bibitem{robinson2014}
P.~A. Robinson, S.~Sarkar, G.~M. Pandejee, and J.~Henderson.
\newblock Determination of effective brain connectivity from functional
  connectivity with application to resting state connectivities.
\newblock {\em Phys. Rev. E}, 90:012707, 2014.

\bibitem{sarkar2013b}
S.~Sarkar, A.~Dong, J.~A. Henderson, and P.~A. Robinson.
\newblock Spectral characterization of hierarchical modularity in product
  architectures.
\newblock {\em Journal of Mechanical Design}, 136:011006, 2013.

\bibitem{sarkar2013a}
S.~Sarkar, J.~A. Henderson, and P.~A. Robinson.
\newblock Spectral characterization of hierarchical network modularity and
  limits of modularity detection.
\newblock {\em PLoS One}, 8(1):e54383, 2013.

\bibitem{sporns2011}
O.~Sporns.
\newblock {\em Networks of the brain}.
\newblock MIT Press, 2011.

\bibitem{sun2009}
L.~Sun, R.~Patel, J.~Liu, K.~Chen, T.~Wu, J.~Li, E.~Reiman, and J.~Ye.
\newblock Mining brain region connectivity for alzheimer's disease study via
  sparse inverse covariance estimation.
\newblock In {\em Proceedings of the 15th ACM SIGKDD Conference}, pages
  1335--1344. ACM, 2009.

\end{thebibliography}
\end{document}